\documentclass [12pt]{iopart}
\usepackage {iopams}
\usepackage{setstack}
\usepackage{amstext}
\usepackage{amsfonts}
\usepackage{graphicx}
\usepackage{subfigure}
\usepackage{array}
\usepackage{appendix}

\usepackage{mathrsfs}
\usepackage{color}
\usepackage{makecell}
\usepackage{adjustbox,lipsum}

\begin{document}

\title[]{Quantum solvability of noisy linear problems by divide-and-conquer strategy}

\author{Wooyeong~Song$^{1,2}$, Youngrong~Lim$^{3}$, Kabgyun~Jeong$^{4,3}$, Yun-Seong~Ji$^{4}$, Jinhyoung~Lee$^{2}$, Jaewan~Kim$^{3}$, M.~S.~Kim$^{5,3}$, and Jeongho~Bang$^{6}$}

\address{$^1$ Center for Quantum Information, Korea Institute of Science and Technology, Seoul, 02792, Korea}
\address{$^2$ Department of Physics, Hanyang University, Seoul 04763, Korea}
\address{$^3$ School of Computational Sciences, Korea Institute for Advanced Study, Seoul 02455, Korea}
\address{$^4$ Research Institute of Mathematics, Seoul National University, Seoul 08826, Korea}
\address{$^5$ QOLS, Blackett Laboratory, Imperial College London, London SW7 2AZ, United Kingdom}
\address{$^6$ Electronics and Telecommunications Research Institute, Daejeon 34129, Korea}

\vspace{10pt}

\begin{indented}
\item The first three authors (W.S., Y.L., and K.J.) contributed equally to this study and can be regarded as the main authors. 
\item Correspondence and requests for materials should be addressed to M.S.K. and J.B.
\end{indented}

\ead{\mailto{m.kim@imperial.ac.uk} and \mailto{jbang@etri.re.kr}}

\begin{abstract}
Noisy linear problems have been studied in various science and engineering disciplines. A class of ``hard'' noisy linear problems can be formulated as follows: Given a matrix $\hat{A}$ and a vector $\mathbf{b}$ constructed using a finite set of samples, a hidden vector or structure involved in $\mathbf{b}$ is obtained by solving a noise-corrupted linear equation $\hat{A}\mathbf{x} \approx \mathbf{b} + \boldsymbol\eta$, where $\boldsymbol\eta$ is a noise vector that cannot be identified. For solving such a noisy linear problem, we consider a quantum algorithm based on a divide-and-conquer strategy, wherein a large core process is divided into smaller subprocesses. The algorithm appropriately reduces both the computational complexities and size of a quantum sample. More specifically, if a quantum computer can access a particular reduced form of the quantum samples, polynomial quantum-sample and time complexities are achieved in the main computation. The size of a quantum sample and its executing system can be reduced, e.g., from exponential to sub-exponential with respect to the problem length, which is better than other results we are aware. We analyse the noise model conditions for such a quantum advantage, and show when the divide-and-conquer strategy can be beneficial for quantum noisy linear problems. 
\end{abstract}


\maketitle

\newtheorem{theorem}{Theorem}

\newcommand{\bra}[1]{\left<#1\right|}
\newcommand{\ket}[1]{\left|#1\right>}
\newcommand{\abs}[1]{\left|#1\right|}
\newcommand{\norm}[1]{\left|\!\left| #1\right|\!\right|}
\newcommand{\expt}[1]{\left<#1\right>}
\newcommand{\braket}[2]{\left<{#1}|{#2}\right>}
\newcommand{\commt}[2]{\left[{#1},{#2}\right]}
\newcommand{\round}[1]{\ensuremath{\lfloor#1\rceil}}

\newcommand{\identity}{1\!\!1}

\section{Introduction}

The number of quantum algorithms that render problems that are prohibitively hard in the classical regime tractable is increasing. In particular, recent research is aimed at developing quantum algorithms that not only reduce the computational complexities but also minimise the resources required for the implementation; and hence, these algorithms can be {\em feasible}. Note that the realisation of such algorithms is becoming a reality even for intermediate-scale (e.g., only a few hundreds of) noisy qubits~\cite{Boixo2018,Bouland2019,Arute2019}.

Owing to their simplicity, linear problems are considered in various practical applications. However, if noise is added, such a problem becomes exponentially difficult, and it can be defined as follows: Given a set of inputs $\mathbf{a}=a_0 a_1 \cdots a_{n-1} \in \mathbb{F}_q^n$ and modularised outputs $b = \mathbf{a} \cdot \mathbf{s} + \eta_\mathbf{a} (\text{mod}~q) \in \mathbb{F}_q$, the ``hidden'' $\mathbf{s}=s_0 s_1 \cdots s_{n-1} \in \mathbb{F}_q^n$ needs to be found in the presence of noise $\eta_\mathbf{a}$ from a distribution $\chi$. Here, $\mathbb{F}_q$ is a finite field of order $q$ and $\mathbb{F}_q^n$ denotes the set of all its $n$ tuples. To avoid confusion, we clarified that the noise $\eta_\mathbf{a}$ is intrinsic to the problem and does not implies any imperfections of a quantum system. In noise-free cases $\eta_\mathbf{a}=0$ ($\forall \mathbf{a}$), $\mathbf{s}$ can be determined, exhibiting only polynomial orders of (classical) samples and time complexities. For example, we can construct a linear equation using a sample set $\left\{ \left( \mathbf{a}_i, b_i=\mathbf{a}_i\cdot\mathbf{s} (\text{mod}~q) \right) \right\}_{i=0}^{M-1}$ ($M \ge n$):
\begin{eqnarray}
\hat{A} \mathbf{x} = \mathbf{b},
\label{eq:linear_eq}
\end{eqnarray}
where $\hat{A}$ is an $n \times n$ matrix whose elements $A_{ij}$ are segments $a_j$ of $\mathbf{a}_i$ in the $i$-th sample and $\mathbf{b}=(b_0, b_1, \ldots, b_{n-1})^T$. Thus, if the equations $\mathbf{a}_i \cdot \mathbf{s} = b_i$ are linearly independent and $\hat{A}^{-1}$ exists, $s_j$ can be obtained for each $j$ from the expression $\mathbf{x} = \hat{A}^{-1}\mathbf{b}$, and $\mathbf{s}$ can be recovered~\cite{Trefethen1997}. Here, $\hat{A}^{-1}$ is a modular inverse matrix. However, noisy and $q$-modularised samples render the task of solving Eq.~(\ref{eq:linear_eq}) very difficult. In particular, the number of samples and time required to obtain a solution $\mathbf{s}$ increase exponentially with the system size $n$~\cite{Regev2009,Regev2010}. The noisy linear problem has been a subject of recent research in quantum computation.

Concurrently, several quantum studies have been conducted to alleviate this noisy linear problem, e.g., by Cross {\em et al.}~\cite{Cross2015} and Grilo {\em et al.}~\cite{Grilo2019}. A common and promising approach is to employ superposed quantum samples, which are defined as
\begin{eqnarray}
\ket{\Psi}=\frac{1}{\sqrt{\abs{V}}}\sum_{\mathbf{a} \in V} \ket{\mathbf{a}}_\mathscr{D} \ket{\mathbf{a} \cdot \mathbf{s} + \eta_\mathbf{a} (\text{mod}~q)}_\mathscr{A},
\label{eq:qn_samples}
\end{eqnarray}
where $V \subseteq \mathbb{F}_q^n$ and $\abs{V}$ is the cardinality of $V$. Here, the number of states involved in the superposition to achieve quantum speedup scales exponentially with $n$, and $\abs{V}$ is of the order, $O(q^n)$\footnote{Throughout this paper, the term ``the size of a quantum sample'' denotes the number of the (classical) input-output pairs $(\mathbf{a}, \mathbf{a} \cdot \mathbf{s} + \eta_\mathbf{a} (\text{mod}~q))$ to be quantum-superposed in constituting $\ket{\psi}$.}. This condition is crucial for achieving exponential speedup and suggests that (A.1) a large [i.e., $O(q^n)$-scaling] set, say $V$, of classical (or non-superposition) samples can be (pre-)allocated in a memory and (A.2) the memory address can be loaded into multiple samples in a superposition. Here, it should be noted that $V$ is not supposed to be directly queried in the quantum setting and exists only for the construction of a quantum sample, as in Eq.~(\ref{eq:qn_samples}). In such a quantum setting, it can be shown that a reasonable quantum speedup can be achieved, requiring only polynomial orders of quantum samples and execution time. 

While Cross {\em et al.} and Grilo {\em et al.} proposed an efficient quantum algorithms for the noisy linear problem, there has been a debate on the total computational complexity, for example, including the preparation of the initial quantum state $\ket{\Psi}$ which is a massively superposed and entangled state. Here, a useful quantum gadget, quantum random access memory (QRAM), can be considered~\cite{Giovannetti2008-PRL,Giovannetti2008-PRA}. However, while the usefulness of QRAM has been outlined, its implementation involves high computational costs~\cite{Paler2020}, and it could affect the speedup achieved in the main computation~\cite{Aaronson2015}. In this study, we consider this very problem. Thus, instead of following the original algorithms which require massive superposition and entanglement, we introduce an approach to ease the tariff to prepare the initial quantum state using the divide-and-conquer strategy. The advantages are straightforward; that is, the algorithm can be executed on the reasonable scale of the quantum gadgets (including QRAM and main computing operations).

In the quantum divide-and-conquer algorithm, the entire system in $\mathbb{F}_q^n$ is divided into $n$ number of segment systems in $\mathbb{F}_q$~\cite{Blum2003, Arora2011}. The debatable issues in such a quantum divide-and-conquer algorithm are: 1) will the divide-and-conquer protocol work in this quantum algorithm, 2) if so, what will be the computational complexity depending on how finely we divide the problem? Answering these questions is not simple. Here, we successfully manage to address the aforementioned issues and prove that even including the state preparation time, we can bring down the computational time to sub-exponential. We also report that the extent to which the size of a quantum sample and its executing system can be reduced is dependent of the noise model and the problem condition. Thus, on the basis of our analysis, we can present a certain criterion for determining whether a reasonable quantum speedup can be achieved when the divide-and-conquer strategy is used.

\section{Resized quantum sample}

Before starting an analysis, the noise model should be defined because it affects the efficiency of an algorithm. First, we consider the case where modulo $q$ of $a_j, b_j \in \mathbb{F}_q$ is sufficiently large and increases with the problem length $n$. Subsequently we consider the noise model, i.e., distribution $\chi$, as being a discrete uniform or bounded Gaussian (frequently referred to as a truncated normal) distribution in the interval $[-\xi, \xi]$ around zero~\cite{Regev2010}. Thus, we have
\begin{eqnarray}
\abs{\eta_{\mathbf{a}}} \le \xi.
\label{eq:b_noise}
\end{eqnarray}
Here, we set $\xi \ll q$. Therefore, the increment in $\xi$ with respect to $n$ is not greater than that in $q$. Such a noise model has been commonly used in relevant studies~\cite{Regev2009,Brakerski2014}.

The divide-and-conquer strategy is used as follows. The system, $\mathbf{a}=a_0 a_1 \cdots a_{n-1} \in \mathbb{F}_q^n$, is divided into $n$ subsystems of scale $\mathbb{F}_q$. Consequently, $n$ $q$-qudits are processed, instead of a $q^n$-qudit. For this strategy, a prior assumption is that we have access to a specific sample in $V$ efficiently to eliminate the $k$-th segment, $a_k$, of another sample. This assumption can be invoked by (A.1) and (A.2). Subsequently, by performing Gaussian elimination(-like) additions of the samples in $V$, e.g., as in the method of ``sample reduction'' in Refs.~\cite{Blum2003,Albrecht2015}, we can construct a resized sample $(a_j', b_j')$ consisting of the following inputs: only the $j$-th element, $a'_j \in \mathbb{F}_q$, and the correct label,
\begin{eqnarray}
b_j' = a_j' s_j + \eta_j',
\end{eqnarray}
which is the $j$-th element of the vector, e.g., $\mathbf{b}'$. Here, the number of sample additions in the Gaussian elimination, denoted as $\kappa$, is at most $O(n^3)$~\cite{Trefethen1997}\footnote{We note that such a quantum scenario is certainly different from the classical one in which the classical oracle is referenced up to $\abs{V}$ times at first to construct $V$, and these oracle calls are considered as the queries to identify $\mathbf{s}$.}. However, it should be noted that the noise spectrum changes from $\eta_{\mathbf{a}}$ to $\eta_j'$ during the sample reduction, because the repetition of the sample additions causes ``spreading out'' of the noise distribution. Thus, the noise bound $\xi$ in Eq.~(\ref{eq:b_noise}) increases, and the increase can be expressed as
\begin{eqnarray}
\xi \to \xi' = \kappa \xi.
\label{eq:chi_change}
\end{eqnarray}
Such a behaviour of the noise in the resized samples limits the applicability of the algorithm. In particular, it strongly depends on the problem condition, i.e., the orders of $q$ and $\xi'$. For example, in the worst case, if $\xi' \gg q$, the noise $\eta_j'$ spreads out uniformly in $[0, q-1]$ because of the modular arithmetic of $b_j' \in \mathbb{F}_q$. In this case, it is impractical to solve the problem. Accordingly, the noise model should be made more specific by assuming the condition,
\begin{eqnarray}
\xi' = \kappa \xi \ll q;
\label{eq:xi'q}
\end{eqnarray}
therefore, the noise distribution $\chi$ maintains its original bounded shape and satisfies
\begin{eqnarray}
\abs{\eta_j'} \le \xi' = \kappa \xi.
\label{eq:b'_noise}
\end{eqnarray}
This leads to the following important condition for our divide-and-conquer strategy to be effective:
\begin{eqnarray}
\text{\em The order of $q$ has to be greater than $O(n^3 \xi)$}.
\label{condi:q}
\end{eqnarray}
\bigskip

Consequently, we can consider a resized quantum sample with the form,
\begin{eqnarray}
\ket{\psi_j} = \frac{1}{\sqrt{\abs{v_j}}} \sum_{a_j' \in v_j} \ket{a_j'}_\mathscr{D'} \ket{a_j' s_j + \eta_j'}_\mathscr{A}, 
\label{eq:q_samples}
\end{eqnarray}
satisfying $\abs{v_j} \le q \ll \abs{V} \le q^n$. Here, we note that the size of the system is reduced, such that $\mathscr{(D,A)} \in \mathbb{F}_q^n \otimes \mathbb{F}_q \to \mathscr{(D', A)} \in \mathbb{F}_q \otimes \mathbb{F}_q$. Assuming that $v_j$ can efficiently be accessed in the superposition, e.g., by implementing QRAM\footnote{Here, we comment that the memory cost can be more saved, such as using the technique in Ref.~\cite{Park2020}.}, we can, in principle, obtain such a quantum sample $\ket{\psi_j}$ in the form of Eq.~(\ref{eq:q_samples}) (see~\ref{appendix:A}). 

\section{Quantum divide-and-conquer algorithm for noisy linear problem}

Here, we propose a quantum algorithm based on the divide-and-conquer strategy. Given $\ket{\psi_j}$ as in Eq.~(\ref{eq:q_samples}), we implement a subroutine of the Bernstein-Vazirani (BV) kernel~\cite{Bernstein1997}, denoted as $\text{BV}(\ket{\psi_j})$. Here, $\text{BV}(\ket{\psi_j})$ consists of two $\text{QFT}_q$'s, each of which is applied to the $j$-th partitioned $\mathscr{D}'$ and $\mathscr{A}$. $\text{QFT}_q$ denotes the $q$-dimensional quantum Fourier transform, 
\begin{eqnarray}
{\text{QFT}_q\ket{j} = \frac{1}{\sqrt{q}}\sum_{k=0}^{q-1}\omega^{jk}\ket{k}},
\end{eqnarray}
where $\omega = e^{i\frac{2 \pi}{q}}$.

After implementing $\text{BV}(\ket{\psi_j})$, we measure the states of $\mathscr{D}'$ and $\mathscr{A}$. We subsequently obtain the $j$-th candidate $\tilde{s}_j$, which is subjected to an $M$-trial test, denoted as ${\cal T}(\tilde{s}_j, M)$, to examine whether $\tilde{s}_j$ is acceptable. If ${\cal T}(\tilde{s}_j, M)$ is completed by accepting $\tilde{s}_j=s_j$, we go on to a $j+1$ . Otherwise, if ${\cal T}(\tilde{s}_j, M)$ fails, then $\text{BV}(\ket{\psi_j})$ is re-implemented with a different $\ket{\psi_j}$ to find and test the other $\tilde{s}_j$ values. These processes---implementing $\text{BV}(\ket{\psi_j})$ and performing ${\cal T}(\tilde{s}_j, M)$---are repeated $L \le q$ times for different candidates $\tilde{s}_j$ until one of them is accepted. Here, the condition $L \le q$ is attributable to the possible number of choices for $\tilde{s}_j$ being $q$. If we accept $\tilde{s}_j$ for each $j$, then $\tilde{s}_0 \tilde{s}_1 \cdots \tilde{s}_{n-1}$ is identified as the solution $\mathbf{s}$. Otherwise, the algorithm retrieves a `failure.' Note that $\mathbf{s}=\tilde{s}_0 \tilde{s}_1 \cdots \tilde{s}_{n-1}$ is dismissed if even if one $\tilde{s}_j$ (among $n$) is not equal to the true segment $s_j$ of $\mathbf{s}$. 

The $M$-trial test ${\cal T}(\tilde{s}_j, M)$ is conducted as follows: First, a `deterministic' (i.e., not superposed or classical) test sample $\ket{t_j}_\mathscr{D'}\ket{t_j s_j + \eta_j'}_\mathscr{A}$ is prepared using the other $n$ original classical samples. Here, $t_j$ is also arbitrarily chosen. Second, after measuring $\ket{t_j}_\mathscr{D'}$ and $\ket{t_j s_j + \eta_j'}_\mathscr{A}$, the following quantity is evaluated:
\begin{eqnarray}
\Delta_j = \abs{(t_j s_j + \eta_j') - t_j \tilde{s}_j} = \abs{t_j (s_j - \tilde{s}_j) + \eta_j'}.
\end{eqnarray}
Note that when $\tilde{s}_j = s_j$, the condition $\Delta_j = \abs{\eta_j'} \le \xi'$ is always true, as suggested by Eq.~(\ref{eq:b'_noise}). Thus, if $\Delta_j \le \xi'$, the aforementioned two steps are performed by preparing other deterministic test samples. When the candidate $\tilde{s}_j$ satisfies $\Delta_j \le \xi'$ for $M$ different test samples, ${\cal T}(\tilde{s}_j, M)$ passes accepting $\tilde{s}_j=s_j$; otherwise, ${\cal T}(\tilde{s}_j, M)$ fails. Here, the probability that $\Delta_j \le \xi'$ is satisfied even when $\tilde{s}_j \neq s_j$ is smaller than $\frac{2\xi' + 1}{q}$. Therefore, the probability that we incorrectly accept a candidate $\tilde{s}_j$ for any $j$ is at most
\begin{eqnarray}
\left( \frac{2\xi' + 1}{q} \right)^M,
\label{eq:prob_test_f}
\end{eqnarray}
which is expected to decay fast because $\xi' \ll q$, as specified in Eq.~(\ref{eq:xi'q}).

\begin{figure}[t]
\centering
\includegraphics[width=1.00\textwidth]{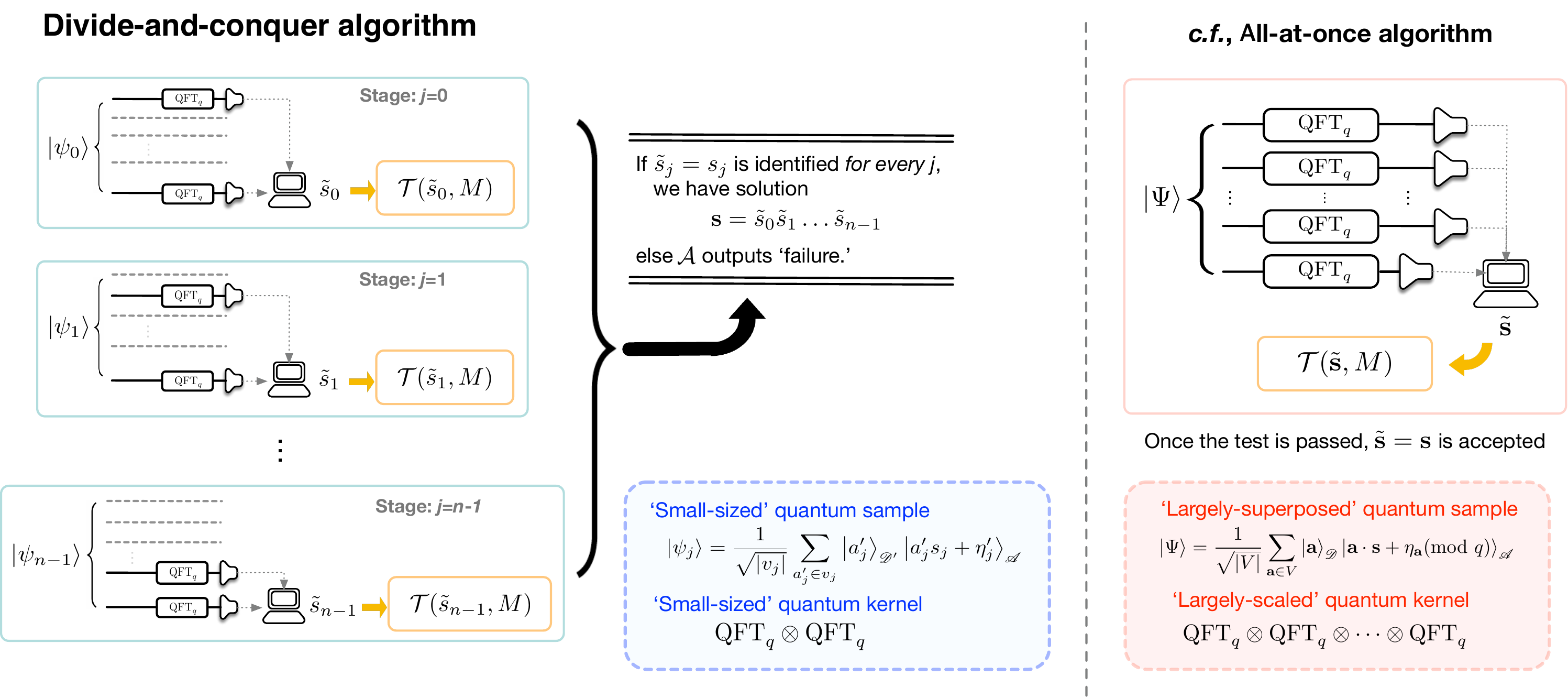}
\caption{\label{fig:algorithm} A schematic of our divide-and-conquer algorithm. The main idea is to divide the system into $n$ subsystems. This enable us to use the `small-sized' learning elements; i.e., the quantum samples $\ket{\psi_j}$ defined as in Eq.~(\ref{eq:q_samples}), and subroutine of the algorithm kernel, denoted by $\text{BV}(\ket{\psi_j})$. In our algorithm, we run the subroutine $\text{BV}(\ket{\psi_j})$ to obtain a candidate fraction $\tilde{s}_j$ for each $j$. With the obtained $\tilde{s}_j$, we also perform a test ${\cal T}(\tilde{s}_j, M)$ by checking if $\tilde{s}_j$ is (in)competent for the true fraction $s_j$. Here, if the test ${\cal T}(\tilde{s}_j, M)$ fails to verify $\tilde{s}_j=s_j$ during $L$ different candidates $\tilde{s}_j$, it returns a null (or empty) output in $j$-th round. In this way, we attempt to find the solution $\mathbf{s}$ by identifying $\tilde{s}_j = s_j$ for every $j$. Here, our algorithm gives ``failure'', if any of $\tilde{s}_0, \tilde{s}_1, \ldots, \tilde{s}_{n-1}$ is not identified with null output (See the main text for details). We also depict all-at-once algorithm for comparison.}
\end{figure}

In Fig.~\ref{fig:algorithm}, the schematic of our divide-and-conquer algorithm is depicted, where all-at-once algorithm is also given for comparison. 

\section{Quantum-sample and time complexities}

Herein, we present the details of the proposed algorithm and the analysis of its computational performance. First, consider the sample state $\ket{\psi_j}$ in Eq.~(\ref{eq:q_samples}). Here, if we assume that there is no noise, i.e., $\eta_j'=0$, then implementing $\text{BV}(\ket{\psi_j})$ yields
\begin{eqnarray}
\frac{1}{q\sqrt{q}} \sum_{a_j' \in \mathbb{F}_q} \sum_{k_j, k^\star \in \mathbb{F}_q} \omega^{a_j' (k_j + s_j k^\star)} \ket{k_j}_\mathscr{D'} \ket{k^\star}_\mathscr{A} = \frac{1}{\sqrt{q}}\sum_{k^\star \in \mathbb{F}_q} \ket{- s_j k^\star}_\mathscr{D'} \ket{k^\star}_\mathscr{A},
\label{eq:output_noerror}
\end{eqnarray}
where we have chosen $v_j=\mathbb{F}_q$ (thus, $\abs{v_j}=q$) and used the discrete $\delta$-function,
\begin{eqnarray}
\delta_{k_j, -\alpha s_j k^\star}=\frac{1}{q} \sum_{a_j' \in \mathbb{F}_q} \omega^{a_j' (k_j + \alpha s_j k^\star)}.
\label{eq:delta}
\end{eqnarray}
Thus, by measuring the states of $\mathscr{D'}$ and $\mathscr{A}$, we can immediately obtain $s_j$ without any test as long as $k^\star \neq 0$. In this case, the probability of identifying $s_j$ is $1-\frac{1}{q}$. Therefore, only $n$ repetitions of the above process are required for finding the solution $\mathbf{s}=s_0 s_1 \cdots s_{n-1}$ without any test. Thus, the total number of quantum samples required to identify $\mathbf{s}$ is approximately $\frac{q}{q-1} n$, and the algorithm accepts polynomial quantum-sample/time complexity. This is comparable to the classical results and there is no quantum speedup.

In a noisy linear problem (i.e., $\eta_j' \neq 0$), after implementing $\text{BV}(\ket{\psi_j})$, the states of $\mathscr{A}$ are not perfectly correlated with those of $\mathscr{D'}$ and we cannot apply the $\delta$-function in Eq.~(\ref{eq:delta}). Instead, states $\ket{k_j}_\mathscr{D'}$ and $\ket{k^\star}_\mathscr{A}$ are correlated as follows:
\begin{eqnarray}
\frac{1}{q \sqrt{\abs{v_j}}} \sum_{a_j' \in v_j}\sum_{k_j, k^\star \in \mathbb{F}_q} \omega^{a_j' (k_j + s_j k^\star) + \eta_j' k^\star} \ket{k_j}_\mathscr{D'} \ket{k^\star}_\mathscr{A}. \nonumber \\
\label{eq:output_LWE}
\end{eqnarray}
Thus, $\text{BV}(\ket{\psi_j})$ yields candidate $\tilde{s}_j$, which is generally not equal to $s_j$. The probability that $\tilde{s}_j$ is equal to $s_j$, denoted as $P(\tilde{s}_j = s_j)$, is calculated by substituting $k_j = - s_j k^\star$ into Eq.~(\ref{eq:output_LWE}) as follows:
\begin{eqnarray}
P(\tilde{s}_j = s_j) &=& \frac{1}{q^2 \abs{v_j}} \norm{ \sum_{k^\star \in \mathbb{F}_q}\sum_{a_j' \in v_j} \omega^{\eta_j' k^\star}\ket{- s_j k^\star}_\mathscr{D}\ket{k^\star}_\mathscr{A} }^2 \nonumber \\
    &\ge& \frac{1}{q^2 \abs{v_j}} \sum_{k^\star \in \mathbb{F}_q} \left( \sum_{a_j' \in v_j} \text{Re}\left(\omega^{\eta_j' k^\star}\right) \right)^2,
\label{eq:P_iden}
\end{eqnarray}
where $\text{Re}(z)$ is the real part of the complex number $z$ and the lower bound in the last line originates from the trivial estimation, $\abs{z}^2 \ge \abs{\text{Re}(z)}^2$. Thus, we can bound the probability $P(\tilde{s}_j = s_j)$ as
\begin{eqnarray}
P(\tilde{s}_j = s_j) \ge \frac{\gamma \abs{v_j}}{\xi' q}\cos^2{\left( 2\pi \gamma\right)}.
\label{eq:lower_b_P}
\end{eqnarray}
This lower bound follows from
\begin{eqnarray}
\text{Re}{\left(\omega^{\eta_j' k^\star}\right)} \ge \cos{\left(2\pi \gamma \frac{\abs{\eta_j'}}{\xi'}\right)} \ge \cos{\left(2\pi \gamma\right)},
\end{eqnarray}
where we let $k^\star \le \round{\frac{\gamma q}{\xi'}}$ with $\gamma \in \left[0, \frac{1}{4} \right)$, which leads to $\sum_{k^\star \in \mathbb{F}_q} \to \sum_{k^\star \le \round{\frac{\gamma q}{\xi'}}}$. 

It remains to be investigated whether the bounds on the quantum-sample/time complexity, i.e., $O(nL)$, reduces to a polynomial in $n$. This can be achieved by analysing the possible results as follows: 
\begin{itemize}
\item[(i)] We can achieve $\mathbf{s} = \tilde{s}_0\tilde{s}_1\cdots\tilde{s}_{n-1}$ by verifying $\tilde{s}_j = s_j$ for each $j$, which indicates successful operation of the algorithm. Here, let the probability of the overall algorithm success be $1-\delta$ (for any $\delta \ge 0$). 
\item[(ii)] The algorithm can return a ``failure'' to identify $\mathbf{s}$ with a null output for any $j$. 
\item[(iii)] Another failure occurs when the algorithm accepts $\tilde{s}_0\tilde{s}_1\cdots\tilde{s}_{n-1}$, even with $\tilde{s}_j \neq s_j$, for any $j$.
\end{itemize}
First, let us analyse case (iii) by considering the probability, denoted as Prob(iii), of incorrectly identifying the solution $\mathbf{s}$. From Eq.~(\ref{eq:prob_test_f}) and by noting that $\tilde{s}_0\tilde{s}_1\cdots\tilde{s}_{n-1} \neq \mathbf{s}$ if even one $\tilde{s}_j$ (among $n$) is not equal to $s_j$, we can bound Prob(iii) as
\begin{eqnarray}
\text{Prob(iii)} \le L \left( \frac{2\xi' + 1}{q} \right)^M.
\label{eq:prob(iii)}
\end{eqnarray}
Here, for large $n$, we expect Prob(iii) to reduce to $0$. Thus, the number, $M$, of test trials should be set such that the order of $\left( \frac{2\xi' + 1}{q} \right)^M$ decreases as fast as $O(L^{-1})$ with an increase in $n$. Thus, once $\tilde{s}_0, \tilde{s}_1, \ldots, \tilde{s}_{n-1}$ are accepted, we can obtain the solution as $\mathbf{s}=\tilde{s}_0 \tilde{s}_1 \cdots \tilde{s}_{n-1}$ confidently.

Nevertheless, the algorithm can return ``failure'' with a certain probability, e.g., Prob(ii). To analyse this, recall Eq.~(\ref{eq:lower_b_P}). Noticeably, for large $n$, the probability of having a null output for any $j$ is at most
\begin{eqnarray}
\left( 1 - C\xi'^{-1} \right)^{L} \simeq \frac{\delta}{n},
\end{eqnarray}
where $C = \gamma\cos^{2}{\left( 2\pi \gamma\right)}$. This approximation can be confirmed by the following critical settings:
\begin{eqnarray}
\abs{v_j}=O(q)~(\forall j)~\text{and}~L=C \xi' \ln{\frac{n}{\delta}}.
\label{eq:critical_condi}
\end{eqnarray}
Here, it is crucial that the order of $q$ in Eq.~(\ref{eq:lower_b_P}) is cancelled by the size of the quantum sample, i.e., $\abs{v_j}$. Note that the condition $L \le q$ should be satisfied. Thus, we can obtain the lower bound of the overall success probability, Prob(i), as follows:
\begin{eqnarray}
\text{Prob(i)} \ge \left( 1 - \frac{\delta}{n} \right)^n \simeq 1-\delta,
\label{eq:prob(i)}
\end{eqnarray}
or, equivalently, the upper bound of the overall failure probability as $\text{Prob(ii)} \le \delta$.

\begin{table*}
\centering
\tabcolsep=0.1in
\begin{adjustbox}{max width=\textwidth}
\begin{tabular}{c c c c}
\hline\hline
Algorithm (Type) & \makecell{Sample Complexity} & \makecell{Time Complexity} & \makecell{Quantum-Sample Size} \\
\hline
\makecell{Blum {\em et al.}~\cite{Blum2003} (classical)} & $2^{O(n/\log{n})}$ & $2^{O(n/\log{n})}$  & --  \\
\makecell{Lyubashevsky~\cite{Lyubashevsky2005} (classical, for $q=2$)} & $n^{1+\varepsilon}$ & $O(2^{n/\log{\log{n}}})$ & --  \\
\makecell{Arora and Ge~\cite{Arora2011} (classical)} & $2^{\tilde{O}(n^{2\varepsilon})}$ (for $\varepsilon < \frac{1}{2}$) & $\Omega(q^2\log{q})$ & --  \\
\makecell{Grilo {\em et al.}~\cite{Grilo2019} (quantum)} & $O(\xi \log{\frac{1}{\delta}})$ (for $\xi \in \text{poly}(n) \ll q$) & $O(\text{poly}(n, \log{\frac{1}{\delta}}))$ & \makecell{$\abs{V}=O(q^n)$} \\
\makecell{Ours (quantum)}     & $O(n \kappa \xi \log{\frac{n}{\delta}})$ (for $\kappa \xi \in \text{poly}(n) \ll q$ with $\kappa = O(n^3)$) & $O(\text{poly}(n, \log{\frac{n}{\delta}}))$ & \makecell{$\abs{v_j}=O(q)$ ($\forall j$)}  \\
\hline\hline
\end{tabular}
\end{adjustbox}
\caption{\label{tab:comparison} Comparison of the complexities of classical and quantum noisy linear algorithms. Note that results of Lyubashevsky's algorithm are case for $q=2$. $\tilde{O}(\cdot)$ is soft-$O$, which is used to ignore logarithm $n$. In classical algorithms, $n^\varepsilon \le \xi$ is considered with $\varepsilon \in (0, 1)$. Here, $1-\delta$ is lower bound of success probability.}
\end{table*}

Finally, we can present the quantum-sample/time complexity of our algorithm. Given the $\xi$-bounded noise distribution $\chi$, the proposed algorithm can learn $\mathbf{s}$ with a probability greater than $1-\delta$. As the total number of quantum samples required to complete the algorithm is at most $n \times L$, the quantum-sample complexity can be expressed as
\begin{eqnarray}
O\left(n \kappa \xi \log{\frac{n}{\delta}}\right)
\label{eq:sample_c}
\end{eqnarray}
using Eq.~(\ref{eq:critical_condi}). Thus, we obtain a polynomial quantum-sample complexity by adopting 
\begin{eqnarray}
\xi = O(\text{poly}(n)).
\label{eq:poly_xi}
\end{eqnarray}
This directly leads to a polynomial time complexity as
\begin{eqnarray}
O\left(\text{poly}(n, \log{\frac{n}{\delta}})\right)
\label{eq:time_c}
\end{eqnarray}
with $\kappa=O(n^3)$. Table~\ref{tab:comparison} presents a comparison of the computational costs of the classical and quantum algorithms.

\section{Discussion}

The quantum sample size, denoted as $\abs{v_j}$, is an important factor for the algorithm. If no superposed sample is used, polynomial quantum-sample and time complexities will not be guaranteed. For instance, let us consider a deterministic sample $\ket{\psi_j} = \ket{a_j'}_\mathscr{D}\ket{a_j' s_j + \eta_j'}_\mathscr{A}$ which is not superposed but still allows to process the quantum parallelism by $\text{QFT}_q$'s in the BV kernel. In this case, the order of $q^{-1}$ in the lower bound of $P(\tilde{s}_j = s_j)$ in Eq.~(\ref{eq:lower_b_P}) cannot be cancelled out because $\abs{v_j}=1$. Thus, we have a condition different from Eq.~(\ref{eq:critical_condi}), and it will affect the computational complexities (a detailed description is provided below).

The divide-and-conquer strategy can allow a smaller degree, i.e., $\abs{v_j} = O(q)$ ($\forall j$), of the quantum sample size compared to $\abs{V}=O(q^n)$. Furthermore, the cost of implementing the $\text{BV}(\ket{\psi_j})$ algorithm kernel is also lowered, requiring only $2 \times O(q \log{q})$ (controlled rotation) operations, compared to those of $(n+1) \times O(q \log{q})$ in Ref.~\cite{Grilo2019}. However, it should be implement $n$ times to complete the algorithm---understandably, operating a $2 \times O(q \log{q})$ circuit $n$ times with the resized quantum sample $\ket{\psi_j}$ [as in Eq.~(\ref{eq:q_samples})] is much easier than operating a large $(n+1) \times O(q \log{q})$ circuit once with the largely superposed one $\ket{\Psi}$ [as in Eq.~(\ref{eq:qn_samples})]. For a more detailed analysis, let us look into the following three specific cases:

{\em Case 1.}---If $q$ is sub-exponential, e.g., $q \in [2^{n^\epsilon}, 2^{1+n^\epsilon})$ for $\abs{\epsilon} < 1$ (similar to the case of a stringent cryptographic scenario~\cite{Brakerski2014} and considered in Ref.~\cite{Grilo2019}), the solution $\mathbf{s}$ can be found in a polynomial time satisfying the critical conditions in Eqs.~(\ref{eq:xi'q}) and~(\ref{eq:poly_xi}). Thus, quantum polynomial solvability is achieved; and the sizes of the quantum samples and the executing system are sub-exponential. Note that such a polynomial solvability is not achievable without a quantum computer, even though a resized quantum sample as in Eq.~(\ref{eq:q_samples}) is available (as argued in Ref.~\cite{Grilo2019}).

{\em Case 2.}---If $q \in \text{poly}(n) \ge O(n^3 \xi)$\footnote{The inequality originates from Eq.~(\ref{condi:q}).}, the reduction in the superposition size from $O(q^n) \to O(q)$ becomes remarkable, i.e., from exponential to polynomial. However, in this case, the cancellation of $O(q^{-1})$ in Eq.~(\ref{eq:lower_b_P}) becomes trivial because the orders of $\xi$ and $q$ are identical. Thus, the classical algorithms can also achieve polynomial sample/time complexity in this case. For example, one may consider a simple strategy which tests each possible number [i.e., $q \in \text{poly}(n)$] of candidates with the non-superposed samples generated by measuring Eq.~(\ref{eq:q_samples}) in the computational basis.

{\em Case 3.}---If we consider another extreme case that $q$ is equal to or larger than the exponential in $n$, polynomial quantum-sample/time complexity can occur and exponential speedup is valid. However, the reduction in the sample (or system) size from $O(q^n)$ to $O(q)$ will be irrelevant and the divide-and-conquer algorithm will not be significantly advantageous over Grilo {\em et al.}'s original algorithm~\cite{Grilo2019}.

The examples in {\em Cases 1--3} indicate when and how polynomial quantum-sample and time complexities can be achieved and to what extent the system size is relevantly reduced. First, as long as we can access the resized sample as in Eq.~(\ref{eq:q_samples}), the divide-and-conquer strategy enables a polynomial quantum-sample/time complexity. However, the reduction in the size of quantum sample, and hence, the system, becomes significant in a particular scaling range of $q$, e.g., as in {\em Case 1}; specifically, when the order of $q$ is larger than that of the optimal classical sample complexities. Otherwise, one can consider the strategy without using a quantum computer, as described in {\em Case 2}, requiring only $\simeq O(nq)$ of the non-superposed samples. For example, if $q$ scales as $2^{O(n/\log{n})}$\footnote{This is an optimal level of sample complexity that can be obtained from the classical algorithm~\cite{Blum2003}.}, one can test ${\cal T}(\tilde{s}_j, M)$ with each candidate $\tilde{s}_j \in [0, q)$ by measuring Eq.~(\ref{eq:q_samples}). In this case, the number of non-superposed samples required to identify $\mathbf{s}$ is approximately $M \times n \times 2^{O(n/\log{n})}$, which is comparable to the result of Ref.~\cite{Blum2003} (for $q=2$), and the quantum computer is not significantly powerful. Conversely, if $q$ scales excessively large, e.g., exponential in $n$, the power of the quantum computer can be confirmed; however, the divide-and-conquer algorithm will not be helpful because the size of the sample $\ket{\psi_j}$ and $\text{BV}(\ket{\psi_j})$ is still exponential.

We have critically analysed  quantum solvability of noisy linear systems, which will place the implication of the achieved speedup in an appropriate context. However, it should be clarified that it is still hard to make a conclusive statement about whether contemporary quantum algorithms are able to overcome the NLP-hardness; and hence, the NLP-based crypto-hardness. There is potential for more efficient realisation of the algorithm with the reduction in the size of the quantum samples.

\section*{Acknowledgements}

J.B. and K.J. are grateful to M. Wie\'{s}niak, M. Paw\l{}owski, M. \.{Z}ukowski, and other members of the International Centre for Theoretical Quantum Technologies (ICTQT) for their helpful discussions. W.S., Y.L., K.J., and J.B. thank the members of the National Security Research Institute (NSRI). W.S. and J.B. thank N. Liu for their valuable discussions and comments. J.B. thanks Y.-S. Kim, Y.-W. Cho, and H.-T. Lim for the discussions on the optics experiments. This study was partly supported by the National Research Foundation of Korea (Nos.~2018R1D1A1B07047512, 2019M3E4A1079666, 2019R1A2C2005504, 2019R1I1A1A01060756, 2020M3E4A1077861, and 2021M3E4A1038213). It was also supported by the Ministry of Science, ICT, and Future Planning by an Institute of Information and Communications Technology Planning and Evaluation grant funded by the Korea government (No.~2020-0-00890, ``Development of trusted node core and interfaces for the interoperability among QKD protocols''). W.S. acknowledge the KIST research program (2E31021). W.S., Y.L. and J.B. acknowledge the research project (No.~2019-100) funded by an ETRI-affiliated research institute. Y.L was supported KIAS Individual Grant (No.~CG073301) at Korea Institute for Advanced Study. J.K. was supported in part by the KIAS Advanced Research Program (No.~CG014604). M.S.K. acknowledges financial supports from the Samsung GRP grant and the EPSRC Quantum Computing and Simulations Hub grant.

\appendix

\section{Schematic of QRAM process to construct Eq.~(\ref{eq:q_samples})}\label{appendix:A}

QRAM can read allocated data output , such that~\cite{Giovannetti2008-PRL}
\begin{eqnarray}
\frac{1}{\sqrt{\abs{R}}}\sum_{{\bf adr} \in R} \ket{\bf adr}\ket{null} \to \frac{1}{\sqrt{\abs{R}}}\sum_{{\bf adr} \in R} \ket{\bf adr}\ket{D_{\bf adr}},
\label{eq:qram_proc}
\end{eqnarray} 
where $\ket{\bf adr}$ is the address state, $\ket{null}$ is the null state, $R$ denotes the space of the addresses, and $D_{\bf adr}$ denotes the data. In our case, $D_{\bf adr}$ can be considered as the resized samples, $(a'_j, a'_j s_j + \eta'_j)_{\bf adr}$, in $v_j$. Note that the samples are allocated by ${\bf adr}$. Thus, letting $\abs{R}=\abs{v_j}$, the address symbol, ${\bf adr}$, can be expressed as a $single$-tuple of binary numbers: ${\bf adr}=r_0 r_1 \ldots r_{l-1}$, where $r_j \in \{0, 1\}$ for all $j=0,1,\ldots, l-1$. Here, $l=\lceil \log_2 \abs{v_j} \rceil$. Thus, Eq.~(\ref{eq:qram_proc}) yields an address-and-sample `entangled' state as
\begin{eqnarray}
\ket{\Psi} = \frac{1}{\sqrt{\abs{R}}}\sum_{{\bf adr} \in R} \ket{\bf adr} \ket{(a'_j, a'_j s_j + \eta'_j)_{\bf adr}}_\mathscr{D',A}
\end{eqnarray}
After decoupling the address and sample (called ``fan-in''), we can obtain $\ket{\psi}$ in the form of Eq.~(\ref{eq:q_samples}). Here, the summation, $\sum_{{\bf adr} \in R}$, can be replaced by $\sum_{a'_j \in v_j}$ because each sample $(a'_j, a'_j s_j + \eta'_j)$ can be matched to its corresponding address ${\bf adr}$. Conceptually, the symbols of ${\bf adr}$ can be ``incorporated into'' or ``synchronised with'' those of $a'_j$.

\section*{References}

\bibliographystyle{iop}

\end{document}